\documentclass[a4paper,11pt]{article}

\usepackage{amsmath, amssymb, a4wide, bm, cancel, color, graphicx, subfig, youngtab, enumerate, appendix}
\usepackage{cite}
\usepackage{hyperref}

\newcommand{\SU}[1]{\mathrm{SU}(#1)}

\newcommand{\U}[1]{\mathrm{U}(#1)}
\newcommand{\Tr}[2]{\mathop{\mathrm{Tr}}_{#1}\left[#2\right]}
\newcommand{\diff}[2]{\frac{{\rm d}#1}{{\rm d}#2}}

\newcommand{\dimint}[2]{\int\mathrm{d}^{#1}#2\,}
\newcommand{\dimintlim}[4]{\int_{#3}^{#4}\mathrm{d}^{#1}#2\,}
\newcommand{\nbrack}[1]{\left(#1\right)}
\newcommand{\sbrack}[1]{\left[#1\right]}

\newcommand{\sep}[1]{\quad\mbox{#1} \quad}
\newcommand{\brm}[1]{\bm{\mathrm{#1}}}
\newcommand{\cbrm}[1]{\overline{\bm{\mathrm{#1}}}}
\newcommand{\vac}[1]{|{\rm vac}\rangle_{#1}}

\def\fund{\tiny\Yvcentermath1\yng(1)}
\def\afund{\tiny\tilde{\Yvcentermath1\yng(1)}}

\def\be{\begin{equation}}
\def\ee{\end{equation}}
\def\ba{\begin{eqnarray}}
\def\ea{\end{eqnarray}}

\def\uno{\mbox{1 \kern-.59em {\rm l}}}

\numberwithin{equation}{section}
\numberwithin{figure}{section}
\numberwithin{table}{section}

\begin{document}

\title{{\normalsize DCPT-10/63\hfill\mbox{}\hfill\mbox{}}\\
\vspace{2.5 cm}
\Large{\textbf{Solitonic supersymmetry restoration}}}
\vspace{2.5 cm}
\author{James Barnard\\[3ex]
\small{\em Department of Mathematical Sciences,}\\
\small{\em Durham University, Durham DH1 3LE, UK}\\[1.5ex] 
\small james.barnard@durham.ac.uk\\[1.5ex]}
\date{}
\maketitle
\vspace{2ex}
\begin{abstract}
\noindent Q-balls are a possible feature of any model with a conserved, global U(1) symmetry and no massless, charged scalars.  It is shown that for a broad class of models of metastable supersymmetry breaking they are extremely influential on the vacuum lifetime and make seemingly viable vacua catastrophically short lived.  A net charge asymmetry is not required as there is often a significant range of parameter space where statistical fluctuations alone are sufficient.  This effect is examined for two supersymmetry breaking scenarios.  It is found that models of minimal gauge mediation (which necessarily have a messenger number U(1)) undergo a rapid, supersymmetry restoring phase transition unless the messenger mass is greater than $10^8$ GeV. Similarly the ISS model, in the context of direct mediation, quickly decays unless the perturbative superpotential coupling is greater than the Standard Model gauge couplings.
\end{abstract}

\section{Introduction}

The problem of how to break supersymmetry (SUSY) is a persistent one and, despite great leaps in understanding, there is yet to be a conclusive solution.  Many of the difficulties arise from the severe restrictions placed on models if they are to avoid having a supersymmetric vacuum.  For example, the Witten index must vanish \cite{Witten:1982df} which requires the model to be either chiral or contain massless matter, both of which pose serious model building difficulties.  In addition, there should be an unbroken U(1)$_R$ symmetry \cite{Nelson:1993nf}.  Not only does this condition greatly restrict the set of available models, it poses some deep phenomenological problems: namely the occurrence of a massless $R$-axion and anomalously small gaugino masses \cite{Komargodski:2009jf}.

An attractive way to sidestep these problems is to abandon the idea of global SUSY breaking and instead allow a supersymmetric vacuum to exist somewhere in the model.  The universe can be placed in a metastable SUSY breaking vacuum and, as long as this vacuum is sufficiently long lived, the aforementioned problems can be ameliorated with no detrimental consequences.  The idea was popularised by Intriligator, Seiberg, and Shih (ISS) in Ref.~\cite{Intriligator:2006dd}, where it was shown that precisely this situation arises naturally in massive SQCD.  There have since been many developments in the area and there are now a plethora of metastable SUSY breaking scenarios to choose from (see Refs.~\cite{Franco:2009wf, Craig:2009hf, Kitano:2010fa, Yanagida:2010wf, Abel:2010uw, Amariti:2010sz, SchaferNameki:2010iz, Craig:2010yf, Behbahani:2010wh, McCullough:2010wf, Abel:2010vb, Brummer:2010zx, Dudas:2010qg} for some recent examples).

One aspect of metastability that has thus far been unexplored is the effect of non-topological solitons, such as Q-balls \cite{Coleman:1985ki}, on the lifetime of the metastable vacuum.  Q-balls exist in many models with scalar fields charged under an unbroken, global U(1) symmetry, and it has long been known \cite{Spector:1987ag, Kusenko:1997hj} that such objects can induce phase transitions.  An important difference between phase transitions precipitated by Q-balls and those arising through more conventional means is that sub-critical vacuum bubbles can build up gradually in the former scenario \cite{Griest:1989bq, Frieman:1989bx, Kusenko:1997hj}: charge conservation ensures stability at any given stage.  The timescale for these decays is therefore much less than one would estimate using the usual tunnelling action.

Many models of metastable SUSY breaking include an unbroken U(1) symmetry in the metastable SUSY breaking vacuum and, of course, support scalar excitations.  It is therefore important to ask whether Q-balls can speed up the decay of this vacuum relative to the usual estimate.  Perhaps of greatest phenomenological significance is the question of whether Q-balls can induce a decay to the true, supersymmetric vacuum, although decays to lower lying SUSY breaking vacua are also important in models with uplifted metastable SUSY breaking vacua \cite{Giveon:2009yu, Abel:2009ze, Koschade:2009qu, Kutasov:2009kb, Barnard:2009ir, Auzzi:2010wm, Maru:2010yx, Curtin:2010ku}.  It transpires that Q-balls have a major effect on the vacuum lifetime and often make seemingly viable metastable vacua decay on a cosmologically negligible timescale.

The structure of this article is as follows.  In section \ref{sec:QB} the salient features of Q-balls and how they induce vacuum decay are reviewed.  It is shown that Q-balls exist in any metastable vacuum with an unbroken U(1) symmetry that does not contain massless, charged scalars.  If there are no charged fermions or vector bosons lighter than the lightest charged scalar and the scalars remain in equilibrium, Q-balls naturally increase in size until they reach a critical charge and the metastable vacuum is destabilised.  Furthermore, statistical charge fluctuations are in themselves often enough to seed the formation of critical Q-balls -- no net charge asymmetry is required.  In sections \ref{sec:GM} and \ref{sec:ISS} these results are applied to a generic model of gauge mediation, and the model of ISS when used in direct mediation.  Both are shown to undergo a premature phase transition to the supersymmetric vacuum for a large range of parameters.  For gauge mediation the vacuum is destabilised unless the messenger mass is greater than $10^8$ GeV.  In the ISS model one must have $h>g_{\rm SM}$.  Section \ref{sec:CON} concludes and discusses some general model building strategies for mitigating the aforementioned effects.

\section{Q-balls in metastable vacua\label{sec:QB}}

Non-topological solitons are common in models whose vacua possess unbroken symmetries.  Global \cite{Coleman:1985ki, Kusenko:1997zq, Enqvist:2003gh}, local \cite{Lee:1988ag}, abelian and non-abelian \cite{Safian:1987pr, Axenides:1998fc} symmetries have been considered, but only the simplest case will be investigated in this article: a global U(1).  Specifically, consider a model of a complex scalar field $\varphi$ in a potential $U(\varphi)$ admitting such a symmetry.  For a vacuum with unbroken U(1) to exist the potential must have a minimum at $\varphi=0$ where, without loss of generality, one can fix $U(\varphi)=0$.  The physical argument for the existence of Q-balls then goes as follows.  Given a total charge $Q$, which is a conserved quantity due to the unbroken U(1) symmetry, one must find the most energy efficient way of distributing it.  If it is more economical to store the charge in a `blob' of non-zero field VEVs than in free scalar particles there are stable Q-ball solutions to the equations of motion.

Mathematically, a field configuration $\varphi(x,t)$ with total charge $Q$ should minimise the energy
\be
E=\dimint{3}{x}\nbrack{\frac{1}{2}|\dot{\varphi}|^2+\frac{1}{2}|\nabla\varphi|^2+U(\varphi)}+
\omega\nbrack{Q-\frac{1}{2i}\dimint{3}{x}\varphi^*\overset{\text{\tiny$\leftrightarrow$}}{\partial}_t\varphi}
\ee
where the second term ensures charge conservation via the Lagrange multiplier $\omega$.  Rearranging gives
\be
E=\dimint{3}{x}\nbrack{\frac{1}{2}|\dot{\varphi}-i\omega\varphi|^2}+
\dimint{3}{x}\nbrack{\frac{1}{2}|\nabla\varphi|^2+U_\omega(\varphi)}+\omega Q
\ee
where an effective potential
\be\label{eq:QBUomega}
U_\omega(\varphi)\equiv U(\varphi)-\frac{1}{2}\omega^2\varphi^2
\ee
has been defined.  All time dependence occurs in the first term from which one deduces that classical solutions take the form $\varphi(x,t)=\varphi(x)e^{i\omega t}$ for some real parameter $\omega$ and a real function $\varphi(x)$ that minimises
\be
\dimint{3}{x}\nbrack{\frac{1}{2}|\nabla\varphi|^2+U_\omega(\varphi)}
\ee
i.e.\ the problem reduces to one of finding the bounce solution associated with tunnelling in three Euclidean dimensions \cite{Coleman:1985ki}.  Solutions to this problem, and therefore Q-balls, generically exist if there are two minima of $U_\omega(\varphi)$ for a finite range of $\omega$; one at $\varphi=0$ and a second \emph{lower} minimum at $\varphi=\varphi_0\neq0$.  When the original potential is everywhere positive this is equivalent to demanding the function $\mu^2(\varphi)=U(\varphi)/\varphi^2$ is minimised away from the origin.

If $U(\varphi)$ is allowed to take negative values the situation changes.  Of interest here are models where the vacuum $\varphi=0$ is only metastable due to the model having a second minimum at $\varphi=\varphi_0\neq0$ where the potential $U(\varphi_0)=U_0$ is negative -- the true vacuum.  The effective potential \eqref{eq:QBUomega} then automatically has a minimum at non-zero $\varphi$ for any choice of $\omega$.  This is clearly true for $\omega=0$ where said minimum coincides with the true vacuum.  As $\omega$ is increased the effective potential becomes more negative and the second minimum is pushed to larger values of $\varphi$ until $\omega=m_\varphi$.  At this point the minimum at the origin is destroyed but, as long as the allowed range of $\omega$ is non-trivial (in other words $m_\varphi>0$) Q-ball solutions will always be supported\footnote{Increasing $\omega$ can never destroy the second minimum as higher order terms protect it.  If they do not the original potential cannot be bounded from below.}.

\begin{figure}[!t]
\begin{center}
\includegraphics[width=0.35\textwidth]{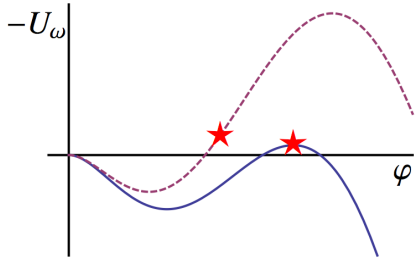}
\caption{{\em When finding Q-ball solutions one can consider damped Newtonian motion of a particle in a potential $-U_\omega(\varphi)$.  The particle comes to rest after infinite time at the origin so it starts just beyond the point where $U_\omega(\varphi)=0$ (stars).  As $\omega$ increases (dashed) the starting point moves closer to the origin and the potential becomes steeper.}\label{fig:bounce}}
\end{center}
\end{figure}

To get a handle on what Q-balls look like one can use their spherical symmetry \cite{Coleman:1985ki} to show that they satisfy
\be
\diff{^2\varphi}{r^2}=-\frac{2}{r}\diff{\varphi}{r}+\diff{U_\omega}{\varphi}
\ee
for radial coordinate $r$.  The problem can be visualised as the damped Newtonian motion of a particle in a potential $-U_\omega(\varphi)$ with respect to `time' $r$.  Q-ball solutions start from rest at non-zero $\varphi$ and come to rest again after infinite time at $\varphi=0$.  In the absence of the damping term the particle would have to start between $\varphi_0$ and zero where $U_\omega(\varphi)=0$, hence the full solution should lie somewhere a little further out than this where $U_\omega(\varphi)<0$.  Since $U_\omega(\varphi)<U(\varphi)$ this alone does not suggest that the actual potential of the Q-ball is negative, but for large enough Q-balls it is indeed the case.  To see why note that decreasing $\omega$ both moves the starting point of the motion away from the origin and decreases the curvature of the effective potential (see Figure \ref{fig:bounce}).  The particle thus starts off more slowly and has further to travel so inevitably takes longer to reach $\varphi=0$.  In terms of the Q-ball this means that $\varphi$ is larger for a greater range of $r$, or small $\omega$ corresponds to a large Q-ball\footnote{See e.g.\ Refs.~\cite{Coleman:1985ki,Kusenko:1997ad} for a more rigourous argument.}.  Conversely, when a Q-ball is large enough $\omega$ is sufficiently small that the interior potential itself is negative, not just the effective potential. 

\begin{figure}[!t]
\begin{center}
\includegraphics[width=0.35\textwidth]{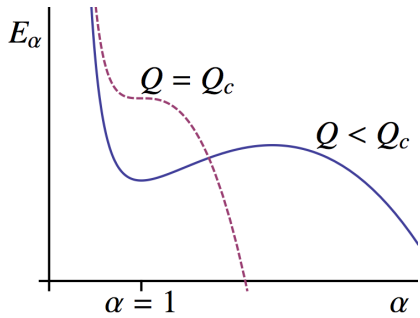}
\caption{{\em Q-ball energy in a metastable vacuum as a function of size $\alpha$.  For charges less than the critical charge $Q_c$ there is a stable solution at $\alpha=1$.  For charges greater than or equal to $Q_c$ there are no stable solutions and the Q-ball expands to fill the universe, initiating a phase transition.}\label{fig:Ealpha}}
\end{center}
\end{figure}

As a function of charge the total Q-ball energy \cite{Coleman:1985ki} is
\be
E(Q)=\frac{Q^2}{2\dimint{3}{x}\varphi(x)^2}+E_S+E_V
\ee
where $E_S$ and $E_V$ are the surface and potential (or volume) energies, and the first term can be thought of as kinetic energy.  The surface energy is always positive\footnote{As discussed in Ref.~\cite{Spector:1987ag} surface energy is critical when the potential is allowed to be negative.  Usually it can be ignored for large Q-balls, but doing so in this case would result in no stable solutions.} but suppose the potential becomes negative and consider varying the radius of the Q-ball by a scaling factor $\alpha$ \cite{Spector:1987ag, Kusenko:1997hj}; the energy goes like
\be\label{eq:SPEQ}
E_\alpha(Q)=\frac{1}{\alpha^3}\frac{Q^2}{2\dimint{3}{x}\varphi(x)^2}+\alpha^2E_S-\alpha^3|E_V|\,.
\ee
For small values of $Q$ there are two stationary points with respect to $\alpha$: a local miniumum at $\alpha=1$ (this is the Q-ball solution so exists by assumption) and a local maximum at $\alpha>1$.  However, as the charge is increased the two solutions move closer together until some critical charge $Q_c$ where there is only one, unstable stationary point.  Above this critical charge $\alpha$ diverges, i.e.\ the Q-ball expands to fill the universe with its own internal state, precipitating a phase transition.  The situation is summarised in Figure \ref{fig:Ealpha}.

To examine the properties of critical Q-balls one typically needs to make some kind of approximation.  First, suppose the Q-ball has radius $R$ and can be approximated by taking $\varphi(r)=\bar{\varphi}=\mbox{constant}$ inside the Q-ball and zero everywhere else.  The energy can be rewritten
\be
E(Q)=\frac{3Q^2}{8\pi R^3\bar{\varphi}^2}+4\pi R^2S-\frac{4}{3}\pi R^3|U(\bar{\varphi})|\,.
\ee
for the surface factor $S$, approximated by \cite{Coleman:1985ki, Kusenko:1997hj}
\be\label{eq:QBS}
S=\dimintlim{}{\varphi}{0}{\bar{\varphi}}\sqrt{2U(\varphi)}\,.
\ee
This is the thin wall limit, and is valid for large Q-balls when the two vacua of the original potential are nearly degenerate -- specifically, the energy barrier $\Delta U$ is much greater than the depth of the global minimum $|U_0|$ (see the solid curve in Figure \ref{fig:bounce}).  One can then take $\bar{\varphi}\approx\varphi_0$ so the Q-ball interior coincides with the true vacuum, upon which the critical charge is defined explicitly by finding the value of $Q$ for which both the first and second derivatives of the energy vanish:
\be
\frac{{\rm d}E}{{\rm d}R}=\frac{{\rm d}^2E}{{\rm d}R^2}=0\quad\implies\quad
Q_c=\frac{100\sqrt{10}\pi S^3\varphi_0}{81|U_0|^{5/2}}\,.
\ee
In addition, one can deduce the value of $R$ for which this occurs and write down the volume and energy of a critical Q-ball
\be\label{eq:QBVcthin}
V_c=\frac{500\pi S^3}{81|U_0|^3}\,,\quad
E(Q_c)=V_c|U_0|
\ee
which will be of use later.

Alternatively the original potential could be such that $\Delta U\lesssim|U_0|$ and the two vacua are highly non-degenerate (the dashed curve in Figure \ref{fig:bounce}).  Now the thick wall approximation \cite{Kusenko:1997ad} is more suitable.  As mentioned earlier a large Q-ball corresponds to small $\omega$ so the limit $\omega\rightarrow0$ can be taken for a critical Q-ball.  The results of Ref.~\cite{Kusenko:1997ad} then yield a critical volume and energy\footnote{Strictly speaking, Ref.~\cite{Kusenko:1997ad} only applies to Q-balls with a small enough charge (or large enough $\omega$).  When $Q$ is too large the energy does not have a stationary point with respect to $\omega$ and the thin wall approximation is used instead.  However, when the potential has a sufficiently deep global minimum the thick wall approximation remains valid even for $\omega\rightarrow0$.  The energy is minimised (albeit is not stationary) at this extremal value of $\omega$.}
\be \label{eq:QBVcthick}
V_c\sim\frac{1}{m_\varphi^3}\,,\quad E(Q_c)\sim V_c\Delta U
\ee
where $\Delta U$ is the height of the potential barrier.  In both thin and thick wall cases the energy approaches a constant value when the charge becomes large.  This is manifest for the thick wall approximation; for the thin wall analogue it can easily be shown using the above equations and noting that $|U_0|$ is small.  Since the critical charge is the point at which the surface and potential energies balance it is perhaps not surprising that an extra unit of charge merits an almost equal and opposite contribution from each.

The key difference between a Q-ball induced phase transition and tunnelling directly to the true vacuum is that a critical Q-ball does {\em not} have to form spontaneously.  Instead it can grow gradually by accreting charge from its surroundings until it reaches the critical size.  Stability is ensured at any stage in this process by charge conservation\footnote{In principle the global U(1) symmetry could be broken in the UV completion of the theory and emerge only as an approximate symmetry in the IR.  This would be the case in stringy models, for example.  The effects of the breaking would be highly suppressed but could provide a decay channel for large Q-balls.  It is not expected that such decay channels would prevent the formation of critical Q-balls but they could slow the process.} and, in models where Q-balls exist, there are classical solutions for arbitrarily small charges \cite{Kusenko:1997ad} (consistent with charge quantisation).  The solutions are expected to be resilient to quantum fluctuations above a charge of about seven \cite{Graham:2001hr}.  In Refs.~\cite{Griest:1989bq, Frieman:1989bx, Kusenko:1997hj} it was shown that one can start from a small Q-ball and gradually build up the charge through ``solitosynthesis'', where the accretion is facilitated by a chain of reactions in thermal equilibrium
\be
\varphi_i+B(Q)\longleftrightarrow B(Q+q_i)
\ee
where $B(Q)$ denotes a Q-ball of charge $Q$.  To reach the critical charge unhindered one requires that the freeze out temperature of the accretion reactions, $T_f$, is less than the temperature at which the critical Q-ball population explodes, $T_c$.  Links in the chain involving Q-balls typically have a large cross section (about the physical size of the Q-ball) so the process is limited by the reactions keeping the $\varphi$'s in equilibrium\footnote{The initial Q-balls can be formed through particle interactions but it is more likely they are remnants of a previous phase transition, or were generated through large field fluctuations when the universe was still very hot.}.  Hence $T_f$ is the freeze out temperature of the $\varphi$'s, whereupon general thermodynamical arguments lead to
\be
T_f\approx\frac{m_{\varphi}}{\ln{(M_{\rm Pl}m_{\varphi}\sigma)}}
\ee
where $\sigma$ is the cross section for the annihilation of $\varphi$'s to light particles.  Meanwhile $T_c$ satisfies \cite{Kusenko:1997hj}
\ba
T_c &=& \nbrack{m_{\varphi}-\left.\diff{E}{Q}\right|_{Q=Q_c}}\nbrack{|\ln{\eta}|+\frac{3}{2}\ln{\frac{m_{\varphi}}{T_c}}-\ln{g_{\varphi}}}^{-1} \nonumber\\
&\approx& \frac{m_\varphi}{|\ln{\eta}|}
\ea
where $\eta$ is the charge asymmetry (the charge per photon) and $g_{\varphi}$ denotes the number of degrees of freedom associated with the $\varphi$'s.  The charge asymmetry is expected to be small so the $|\ln{\eta}|$ term dominates the denominator whereas the derivative term has been set to zero owing to the reasons presented earlier.

Enforcing the inequality $T_c>T_f$ thus collapses to a bound on the charge asymmetry
\be\label{eq:QBetalim}
\eta>\frac{1}{M_{\rm Pl}m_\varphi\sigma}\,.
\ee
Even if the overall charge of the universe is zero, there will always be a statistical contribution to $\eta$ over a finite region of space.  In a comoving volume $V$, the relative excess of charge goes like $1/\sqrt{n_\varphi V}$ ($n_\varphi V$ being the total number of $\varphi$'s) so the charge asymmetry is given by
\be
\eta_{\rm stat}(V)=\frac{n_\varphi-n_{\varphi^*}}{n_\gamma}=\nbrack{\frac{n_\varphi-n_{\varphi^*}}{n_\varphi}}\frac{n_\varphi}{n_\gamma}\sim\frac{1}{\sqrt{n_\varphi V}}\frac{n_\varphi}{n_\gamma}
\ee
for messenger and photon number densities $n_\varphi$ and $n_\gamma$.  From Eq.~\eqref{eq:QBetalim} this is large enough to support critical Q-balls all the way down to the freeze out temperature in any volume
\be
V<\frac{n_\varphi}{n_\gamma^2}(M_{\rm Pl}m_\varphi\sigma)^2\sim
\frac{M_{\rm Pl}\sigma}{m_\varphi^2}\sbrack{\ln{\nbrack{M_{\rm Pl}m_\varphi\sigma}}}^{9/2}\,.
\ee
If a critical Q-ball comfortably fits into this volume, i.e.\
\be\label{eq:QBVclim}
V_c\ll\frac{M_{\rm Pl}\sigma}{m_\varphi^2}\sbrack{\ln{\nbrack{M_{\rm Pl}m_\varphi\sigma}}}^{9/2}
\ee
statistical fluctuations alone are enough to seed their formation.  Physically, one expects to form equal numbers of critical Q-balls carrying both positive and negative charge by this approach.  As they destabilise and expand, their boundaries inevitably collide and the net charge is annihilated when the system relaxes to the true vacuum.

The final obstacle to solitosynthesis is the presence of light, charged fermions or vector bosons.  If these exist, they will absorb any charge instead of the Q-balls and solitosynthesis will not take place.  Since all charged scalars are automatically massive enough not to encroach on critical Q-ball formation, it is sufficient to check that none of the charged fermions and vector bosons in equilibrium at $T_c$ are lighter than the lightest charged scalar.  Assuming this and all other conditions are met, the time at which the phase transition happens can be estimated via the Hubble time scale at the critical temperature:
\be\label{eq:QBtc}
t_c\sim\frac{M_{\rm Pl}}{T_c^2}\sim\frac{M_{\rm Pl}}{m_\varphi^2}\sbrack{\ln{\nbrack{M_{\rm Pl}m_\varphi\sigma}}}^2
\ee
for the minimal required charge asymmetry \eqref{eq:QBetalim}.  Ignoring the log term, this would demand a scalar mass less than $10^{-11}$ GeV for $t_c$ to be less than the age of the universe, $10^{10}$ years.

\begin{figure}[!t]
\begin{center}
\includegraphics[width=0.35\textwidth]{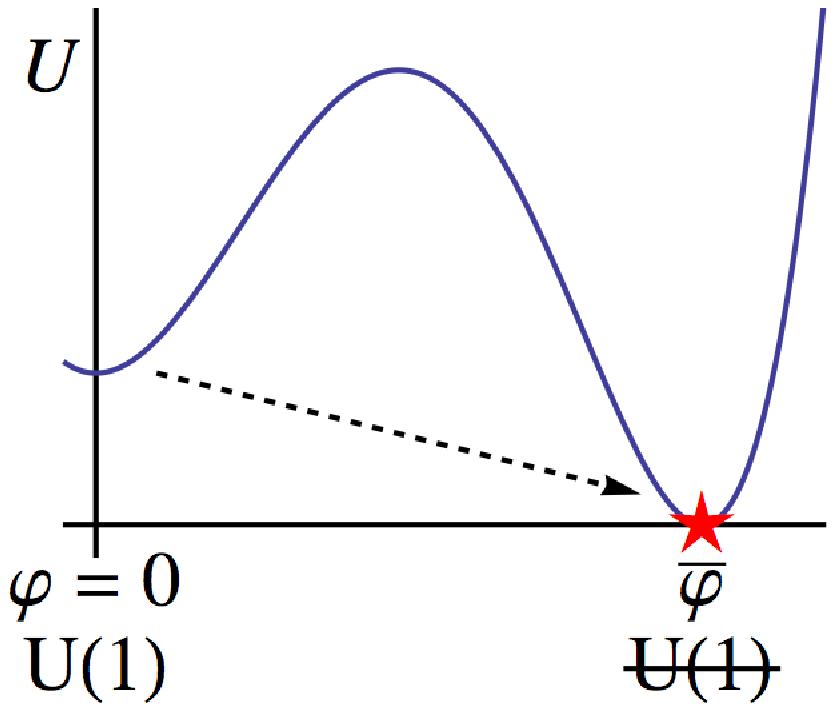}\hspace{20mm}
\includegraphics[width=0.35\textwidth]{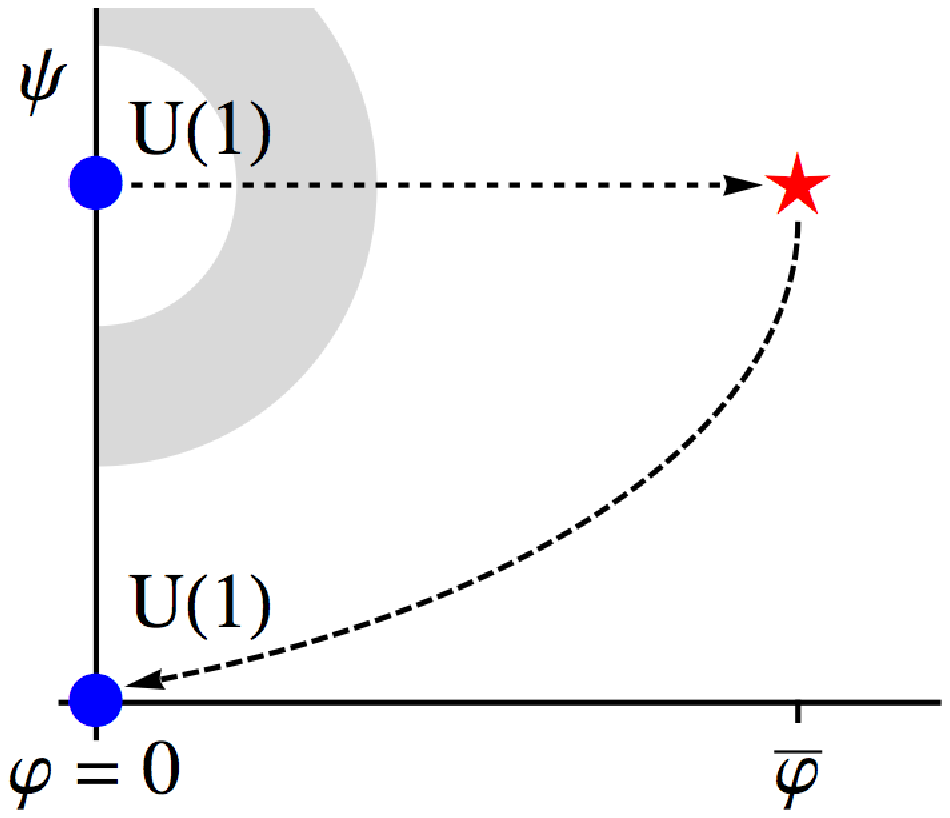}
\caption{Left: {\em Direct decay.  The interior state of the Q-ball (star) coincides with the true vacuum with broken U(1).  Decay from the metastable vacuum proceeds directly through synthesis of critically charged Q-balls}.  Right: {\em Indirect decay.  The metastable vacuum (top circle) decays through a potential barrier (shaded) through synthesis of critically charged Q-balls involving the field $\psi$.  After this decay, the model flows to the true vacuum (bottom circle) where U(1) is restored.}\label{fig:qdecay}}
\end{center}
\end{figure}

When they do occur, phase transitions can occur in one of two ways as illustrated in Figure \ref{fig:qdecay}.  First is {\em direct} decay: the metastable vacuum decays, via Q-balls, directly to the true vacuum.  In this approach the interior of the Q-ball is stable in all directions and roughly coincides with the true vacuum, where the corresponding U(1) symmetry is necessarily broken.  A more subtle variation that can occur when the model contains multiple fields is {\em indirect} decay.  Even if the U(1) symmetry persists in the true vacuum there is always a finite region around it where the symmetry is broken yet the potential remains negative.  Such a region can support Q-ball solutions that quickly decay to the true vacuum once the phase transition has taken place.  In other words, Q-balls can do the hard work of tunnelling through the potential barrier after which the model quickly completes the transition of its own accord.

In summary, Q-balls generically lead to vacuum decay in models of massive scalar fields where the metastable vacuum supports an unbroken, global U(1) symmetry.  The main barriers to decay are cosmological: either the reactions responsible freeze out before critical Q-balls become important or light fermions/vector boson absorb charge in place of Q-balls.  The former does not occur if the charge asymmetry satisfies Eq.~\eqref{eq:QBetalim}, the latter if any charged fermions or vector bosons are heavier than the lightest charged scalar.  Furthermore if Eq.~\eqref{eq:QBVclim} is true, statistical fluctuations in the charge asymmetry can seed critical Q-balls.  Regardless, the lifetime of the metastable vacuum is given by Eq.~\eqref{eq:QBtc}.  These results are readily generalised to models with multiple scalar fields using Ref.~\cite{Kusenko:1997zq}.

\section{Messenger number and gauge mediation\label{sec:GM}}

It is now possible to apply these ideas to some popular models of metastable SUSY breaking and search for instances of solitonic SUSY restoration (SSR).  A simple example of direct SSR can be found in the minimal model of gauge mediation \cite{Giudice:1998bp}.  SUSY is broken by the $F$-term of a chiral superfield $X$ acquiring a VEV, then transmitted to the visible sector by a single pair of messenger chiral superfields $\varphi$ and $\tilde{\varphi}$ charged under the Standard Model gauge group (e.g.\ in the $\brm{5}+\cbrm{5}$ representation of SU(5)).  Communication is via superpotential interactions
\be\label{eq:GMW}
W=W_{\rm SB}(X,\psi)+X\tilde{\varphi}\varphi+M\tilde{\varphi}\varphi\,.
\ee
$W_{\rm SB}$ denotes the (unspecified) superpotential of the SUSY breaking sector -- it is a function of $X$ and some other chiral superfields $\psi$ -- and an explicit messenger mass $M$ has also been included (the coupling constant in the second term has been absorbed into the field $X$).  This model clearly admits a global U(1) messenger number symmetry, under which $\varphi$ and $\tilde{\varphi}$ have charges $+1$ and $-1$.  Messenger number is in fact necessarily conserved in any vacuum that does not break the Standard Model gauge group.  Note that the explicit mass term precludes the possibility of the model having an $R$-symmetry.  This is not strictly necessary, but allows for large gaugino masses due to the model having a supersymmetric vacuum \cite{Nelson:1993nf}.

The $F$-terms derived from Eq.~\eqref{eq:GMW} are
\ba
F_X &=& F(X,\psi)+\tilde{\varphi}\varphi \nonumber\\
F_\varphi &=& (M+X)\tilde{\varphi} \nonumber\\
F_{\tilde{\varphi}} &=& (M+X)\varphi
\ea
where $F(X,\psi)=\partial W_{\rm SB}/\partial X$ is responsible for breaking SUSY in the metastable vacuum, where it takes the value $F$.  The remaining $F$-terms are set to zero by choosing
\be
\vac{+}:\quad\quad\varphi_+=0\,,\quad\tilde{\varphi}_+=0
\ee
whereas the VEV of $X$, fixed at some value $X_+$, is determined by the details of the SUSY breaking sector.  It is straightforward to check that the tree level messenger masses are
\be\label{eq:GGmasses}
m^2_0=\bar{M}^2\pm F\,,\quad
m^2_{1/2}=\bar{M}^2\sep{where}\bar{M}=|M+X_+|
\ee
with the usual requirement $\bar{M}^2>|F|$ imposed by vacuum stability.  $X$ is the goldstino superfield (models where the goldstino is a linear combination involving other fields will be discussed later) so its scalar component is a classical flat direction, or pseudo-modulus \cite{Ray:2006wk, Komargodski:2009jf}.  Hence the VEV of $X$ can be chosen at will without affecting the tree level potential and one finds an additional, supersymmetric minimum at
\be
\vac{0}:\quad\quad\tilde{\varphi}_0\varphi_0=-F\,,\quad X_0=-M\,,\quad U_0=-F^2\,.
\ee
Of course, for the SUSY breaking vacuum to be locally stable the scalar component of $X$ must be stabilised by a mass term arising from loop corrections.  This mass term disappears in the supersymmetric vacuum where the loop correction vanish.

The above model has all the ingredients for SSR; a metastable vacuum with an unbroken U(1) symmetry, charged scalar fields and a supersymmetric vacuum where the symmetry is broken.  However, before formally reaching any conclusion it must be checked that Q-balls can be built up via solitosynthesis (i.e.\ there are no light charged fermions or vector bosons and the messengers remain in thermal equilibrium) and that the critical Q-ball formation temperature $T_c$ is greater than the freeze out temperature of the messengers $T_f$ (i.e.\ Eq.~\eqref{eq:QBetalim} is satisfied).  The only fields carrying messenger number are the messengers themselves and their masses were calculated in Eq.~\eqref{eq:GGmasses}: there is always a charged scalar lighter than the lightest charged fermion so solitosynthesis proceeds unhindered.  Meanwhile messengers are kept in thermal equilibrium by Standard Model gauge interactions, at the messenger scale $\bar{M}$, so the appropriate freeze out cross section is $\sigma\sim g_{\rm SM}^2/\bar{M}^2\sim1/\bar{M}^2$.  One thus finds that any charge asymmetry
\be\label{eq:GGeta}
\eta\gtrsim\frac{\bar{M}}{M_{\rm Pl}}
\ee
is enough to build critical Q-balls.

Statistical fluctuations of $\eta$ alone can seed critical Q-ball formation if Eq.~\eqref{eq:QBVclim} is satisfied, but there are two cases one must consider.  If the SUSY breaking is small then the messenger mass $\bar{M}$ is much greater than the supersymmetric mass splitting $\sqrt{F}$, and therefore the potential barrier is much larger than the difference in energy between the metastable and supersymmetric vacua.  This is the thin wall limit so the critical volume is given by Eq.~\eqref{eq:QBVcthin}:
\be
V_c\approx50\nbrack{\frac{\bar{M}}{F}}^3
\ee
where the surface factor \eqref{eq:QBS} has been approximated in the limit $F\ll\bar{M}^2$ by
\be
S\approx\sqrt{(X_0-X_+)^2+2(\varphi_0-\varphi_+)^2}\sqrt{2U_0}\approx\sqrt{2}\bar{M}F\,.
\ee
Eq.~\eqref{eq:QBVclim} is satisfied unless
\be
\bar{M}>\nbrack{\frac{F}{\bar{M}}}^{3/4}\nbrack{\frac{M_{\rm Pl}}{50}}^{1/4}\sbrack{\ln{\nbrack{\frac{M_{\rm Pl}}{\bar{M}}}}}^{9/8}\,.
\ee
leaving open a substantial region in parameter space for which all constraints for SSR are satisfied and the metastable vacuum decays.  Conversely, if SUSY breaking is large $F$ and $\bar{M}^2$ are of a similar order and one should use the thick wall limit \eqref{eq:QBVcthick} for the critical volume instead. The result is a modified bound
\be
\bar{M}>M_{\rm Pl}\sbrack{\ln{\nbrack{\frac{M_{\rm Pl}}{\bar{M}}}}}^{9/2}
\ee
for a viable vacuum, that cannot hold for any messenger mass below the Planck scale.  In other words, SSR always takes place in the case of large SUSY breaking.

To summarise, Q-balls lead to metastable vacuum decay in any model of minimal gauge mediation unless
\be
\bar{M}>\nbrack{\frac{F}{\bar{M}}}^{3/4}\nbrack{\frac{M_{\rm Pl}}{50}}^{1/4}\sbrack{\ln{\nbrack{\frac{M_{\rm Pl}}{\bar{M}}}}}^{9/8}
\ee
where $\bar{M}$ is the messenger mass and $F$ the scale of SUSY breaking.  If this inequality is not satisfied critical Q-balls are seeded by statistical fluctuations in the charge asymmetry and go on to destabilise the SUSY breaking vacuum.  Using the fact that the gaugino mass is given by $F/16\pi^2\bar{M}\sim1$ TeV, one finds an absolute bound
\be\label{eq:GMfincons}
\bar{M}>10^8\mbox{ GeV}
\ee
on the messenger mass.  When SSR does occur, is does so at time \eqref{eq:QBtc}
\be
t_c\sim\frac{M_{\rm Pl}}{\bar{M}^2}\sbrack{\ln{\nbrack{\frac{M_{\rm Pl}}{\bar{M}}}}}^2
\ee
and is much less than the age of the universe, $10^{10}$ years, for all realistic choices of messenger mass (e.g.\ $t_c\sim10^{-8}$ s for $\bar{M}\sim1$ TeV and decreases as $\bar{M}$ gets larger).

\subsection{Beyond minimal gauge mediation}

In the above only a stripped down version of gauge mediation was considered.  The most obvious way to go beyond the minimal model is to add more messenger fields.  This barely changes the conclusions reached above.  Indeed, one extends the superpotential \eqref{eq:GMW} to
\be
W=W_{\rm SB}(X,\psi)+\lambda_{ij}X\tilde{\varphi}_i\varphi_j+M_{ij}\tilde{\varphi}_i\varphi_j
\ee
for some coupling constants $\lambda_{ij}$.  The messenger mass matrix $\lambda X_++M$ can always diagonalised, upon which the constraints \eqref{eq:GMfincons} must be satisfied with $\bar{M}$ replaced by min$(|\lambda X_++M|)$.

Alternatively one could generalise the SUSY breaking sector so $X$ does not coincide exactly with the goldstino superfield, hence is not necessarily a pseudo-modulus.  One must then consider the details of the SUSY breaking sector to see if and where a supersymmetric vacuum occurs.  Assuming there is one, SSR proceeds much as before.  Light charged fermions and vector bosons remain absent and the temperatures $T_f$ and $T_c$ depend only on the messenger sector so are unchanged.  Even if messenger number persists in the new supersymmetric vacuum there will be a finite region around it where the messenger VEV is non-zero and the relative potential is negative, allowing for indirect SSR.  However, the final constraints \eqref{eq:GMfincons} will be different; they depend on how fields from the SUSY breaking sector affect the Q-ball configuration.  For example, new field VEVs appear in the surface factor \eqref{eq:QBS} and subsequently the critical volume.  This in turn is vital for figuring out whether statistical charge fluctuations can seed critical Q-balls, which must be considered on a case by case basis.

\section{Baryon number and the ISS model\label{sec:ISS}}

One may also ask whether Q-balls can destabilise metastable vacua in the absence of an explicit messenger sector.  A popular model of metastable SUSY breaking is that of Intriligator, Seiberg and Shih (ISS) \cite{Intriligator:2006dd}.  To recap, this model is based on SQCD with gauge group SU($N$), $N_f$ flavours of quark chiral superfield $q$ and $\tilde{q}$, and a gauge singlet meson chiral superfield $\Phi$.  It arises dynamically through Seiberg duality \cite{Seiberg:1994pq} as the low energy limit of massive SQCD with $N_c=N_f-N$ colours and no meson superfield.  At tree level the model has superpotential
\be
W_{\rm tree}=h\tilde{q}\Phi q-hm^2\Tr{}{\Phi}
\ee
for some perturbative coupling $h$ and mass scale $m$ (both assumed real for simplicity), and permits a global symmetry group $\SU{N_f}\times\U{1}_B\times\U{1}_R$ under which
\ba
q &\in& (\fund,\,+1,\,0) \nonumber\\
\tilde{q} &\in& (\afund,\,-1,\,0) \nonumber\\
\Phi &\in& (\brm{Adj+1},\,0,\,2)\,.
\ea
For $N_f>3N$ the model is infrared free and supersymmetry is broken due to the rank condition: the $F$-terms $F_\Phi=h\tilde{q}q-hm^2\uno_{N_f}$ cannot all be satisfied simultaneously as $\tilde{q}q$ is a rank $N$ matrix, $\uno_{N_f}$ is a rank $N_f$ matrix and $N_f>N$.  The metastable vacuum is defined by
\be
\vac{+}:\quad\quad\tilde{q}_+q_+=m^2{\rm diag}(\uno_{N},\,0)\,,\quad\Phi_+=0
\ee
the vacuum energy is $(N_f-N)h^2m^4$ and there is a residual global symmetry group of $\SU{N}\times\SU{N_f-N}\times\U{1}_{B^\prime}\times\U{1}_R$.  Moving the meson VEV away from the origin results in the quarks picking up a tree level mass $h\Phi$.  When this mass becomes large enough the quarks are integrated out, leaving a theory of mesons with a dynamically generated superpotential
\be\label{eq:ISSWdyn}
W_{\rm dyn}=N\nbrack{h^{N_f}\Lambda^{(3N-N_f)/N}\det{\Phi}}^{1/N}-hm^2\Tr{}{\Phi}
\ee
for a dynamical scale $\Lambda$.  One thus finds a supersymmetric minimum at
\be
\vac{0}:\quad\quad\tilde{q}_0=0\,,\quad q_0=0\,,\quad \Phi_0=\frac{1}{h}\Lambda\nbrack{\frac{m}{\Lambda}}^{2N/(N_f-N)}\uno_{N_f}\,.
\ee
As long as $m\ll\Lambda$ all calculations are under control and the metastable vacuum is seemingly long lived.

The U(1) symmetries in the SUSY breaking vacuum comprise a baryon number and an $R$-symmetry, however the $R$-symmetry is anomalous\footnote{Besides, spontaneously broken $R$-symmetries come with an exactly massless, charged fermion in SUSY breaking vacua -- the goldstino -- so do not allow for solitosynthesis.  This renders them defunct from an SSR point of view.} leaving baryon number as the prime candidate for SSR.  It is convenient to expand around the metastable vacuum using degrees of freedom
\begin{align}
\Phi&=\nbrack{\begin{array}{cc} Y & \tilde{Z} \\ Z & X \end{array}} &
q&=\nbrack{\begin{array}{c} m\uno_N+\chi \\ \rho \end{array}} &
\tilde{q}^T&=\nbrack{\begin{array}{c} m\uno_N+\tilde{\chi} \\ \tilde{\rho} \end{array}}
\end{align}
($Y$ and $X$ are $N\times N$ and $(N_f-N)\times(N_f-N)$ matrices and the dimensions of the other components follow suit) with baryon numbers
\ba
B^\prime(Y)=B^\prime(X)=B^\prime(\chi)=B^\prime(\tilde{\chi}) &=& 0 \nonumber\\
B^\prime(\rho)=B^\prime(\tilde{Z}) &=& +1 \nonumber\\
B^\prime(\tilde{\rho})=B^\prime(Z) &=& -1\,.
\ea
The bosonic and fermionic components of all charged fields acquire masses of order $hm$, with the exception of the scalar combinations $\Re[\rho+\tilde{\rho}]$ and $\Im[\rho-\tilde{\rho}]$ which are massless Goldstone bosons of the various broken symmetries.

The presence of these massless, charged scalars prevents the formation of Q-balls.  To see why, note that the effective potential \eqref{eq:QBUomega} has no minimum at the origin for any non-zero value of $\omega$: the addition of a mass term $-1/2\omega^2\varphi^2$ to all charged scalars renders any massless ones tachyonic.  However, when the ISS sector is employed in a direct mediation scenario (as in often the case \cite{Giveon:2009yu, Koschade:2009qu, Barnard:2009ir, Auzzi:2010wm, Maru:2010yx, Curtin:2010ku, Kitano:2006xg, Csaki:2006wi, Abel:2007jx, Haba:2007rj, Zur:2008zg, Abel:2008tx, Abel:2008gv, Essig:2008kz}) the flavour group is gauged; the prospective Goldstone bosons are eaten by gauge fields of the broken symmetry which gain a mass $g_{\rm SM}m$ via the super Higgs mechanism.  Another subtlety arises from the fact that baryon number remains unbroken in the supersymmetric vacuum.  This is not actually a problem as there must exist a finite region in field space around the supersymmetric vacuum where baryon number \emph{is} broken but the potential remains negative.  Indirect SSR remains possible.  Consider, for example, starting from the supersymmetric vacuum and giving all components of the meson a VEV $\hat{\Phi}$, but leaving the quark VEVs fixed at zero.  The relative scalar potential is calculated from Eq.~\eqref{eq:ISSWdyn} and goes like
\ba
U(\hat{\Phi}) &\sim& Nh^2m^4-2N_fh^{(N_f+N)/N}m^2\Lambda^{(3N-N_f)/N}\hat{\Phi}^{(N_f-N)/N}+
\nonumber\\&& N_fh^{2N_f/N}\Lambda^{2(3N-N_f)/N}\hat{\Phi}^{2(N_f-N)/N}\,.
\ea
It is negative for a range of $\hat{\Phi}$ with similar magnitudes to $\Phi_0$, and baryon number is broken whenever the $Z$'s are non-zero.  Hence Q-ball solutions will occur along $Z$ directions with interior values of order $\Phi_0$.

Solitosynthesis is unimpeded if $h\le g_{\rm SM}$ such that the charged gauge bosons are not lighter than their scalar counterparts.  All charged fermions already satisfy this constraint as their masses are equal to those of the scalars at $hm$.  The only thing left to check is whether Eq.~\eqref{eq:QBVclim} holds, such that statistical fluctuations in the charge asymmetry are capable of seeding critical Q-balls.  One must first decide whether to work in the thin or thick wall limit for the purposes of calculating the critical volume.  Since the charged scalars have masses $hm$ and the metastable vacuum energy is order $h^2m^4>(hm)^4$ (recall, $h$ is a perturbative coupling constant) the vacua are highly non-degenerate and the thick wall approximation is most suitable.  Once again the scalars are kept in equilibrium through Standard Model gauge interactions so $\sigma\sim g_{\rm SM}^2/h^2m^2\sim1/h^2m^2$ and, using the critical volume given in Eq.~\eqref{eq:QBVcthick} with $m_\varphi=hm$, Eq.~\eqref{eq:QBVclim} is satisfied unless
\be\label{eq:ISSfincons}
hm>M_{\rm Pl}\sbrack{\ln{\nbrack{\frac{M_{\rm Pl}}{hm}}}}^{9/2}\,.
\ee
The result is, in fact, identical to that found for large SUSY breaking in minimal gauge mediation but with a messenger mass $hm$.  This is of course due to the fact that direct mediation in the ISS model is a specific example of gauge mediation -- the $\rho$'s and $Z$'s act as messengers and their masses are $hm$.  The conclusions reached in the previous section thus hold here; any realistic choice of $hm$ (i.e.\ less than about $M_{\rm Pl}$) results in SSR on a timescale much less than the age of the universe.

In summary, SSR occurs in models of direct mediation using an ISS SUSY breaking sector, unless $h>g_{\rm SM}$ where $g_{\rm SM}$ is the Standard Model gauge coupling and $h$ a perturbative coupling constant.  It does not occur when the Goldstone bosons in the metastable vacuum corresponding to broken flavour symmetries remain massless, but any mechanism or deformation that gives them a mass would yield similar results.  The phase transition takes place at time \eqref{eq:QBtc}
\be
t_c\sim\frac{M_{\rm Pl}}{h^2m^2}\sbrack{\ln{\nbrack{\frac{M_{\rm Pl}}{hm}}}}^2
\ee
which is again much less than the age of the universe unless $hm$ is extremely small (and well into the observable range).

\section{Conclusions\label{sec:CON}}

Q-balls induce vacuum decay in a large class of models, and those of metastable SUSY breaking are no exception.  So long as the metastable vacuum has an unbroken, global U(1) symmetry and no massless, charged scalars Q-balls always exist, with negative interior potential for a sufficiently large charge.  After reaching a critical charge, they precipitate a phase transition to the supersymmetric vacuum regardless of whether or not the U(1) symmetry is preserved there.  In the absence of light charged fermions or vector bosons, and if the scalars remain in thermal equilibrium, these critical Q-balls are able to build up gradually via solitosynthesis and are stable throughout the process due to charge conservation.  Furthermore, there is a range of parameter space where critical Q-balls can be seeded by statistical charge fluctuations alone so no overall charge asymmetry is required.  In order to calculate this range one must decide whether to work in the thick or thin wall approximation, depending on whether the scale of SUSY breaking is comparable to or much less than the charged scalar masses respectively.  Having done so, the timescale of vacuum decay can easily be estimated and is almost always much less than the age of the universe.

These ideas have been applied to models of gauge mediation, where messenger number plays the role of the global U(1) symmetry, and direct mediation in the ISS model, where one can use baryon number.  In both cases, solitonic SUSY restoration occurs.  Any messenger mass less than $10^8$ GeV results in vacuum decay with no net charge asymmetry in the minimal model of gauge mediation, with non-minimal models yielding similar (but model dependent) constraints.  On the other hand messengers lighter than about 3 TeV are expected to overclose the universe, presenting a serious model building challenge.  In the ISS model, SSR is dependent on the Goldstone bosons charged under baryon number getting a large enough mass.  Direct mediation results in them being eaten by gauge bosons, that acquire a mass proportional to the Standard Model gauge couplings.  Unless the gauge couplings are smaller than the perturbative coupling in the superpotential ($h>g_{\rm SM}$) solitosynthesis, ergo SSR, are unimpeded.

The simplest way to build long lived models of metastable SUSY breaking is to ensure that the spectrum contains massless charged scalars or charged fermions lighter than the lightest charged scalar.  $R$-symmetries, for example, always meet this condition due to the massless Goldstino that is associated with SUSY breaking.  Meanwhile the vanilla ISS model is also safe, as there are massless scalars with non-zero baryon number in the metastable vacuum which prevent the formation of Q-balls.  Alternatively one could search for models of SUSY breaking with no unbroken U(1) symmetries at all, although even then one would have to check whether non-topological solitons corresponding to non-abelian symmetries resulted in SSR.  Otherwise SSR is avoided by ensuring that there is no net charge asymmetry in the universe, and that statistical charge fluctuations are insufficient to build critical Q-balls.  It should be noted that the analysis assumes a reheat temperature greater than $T_f\sim m_\varphi$.  Another way to evade these conclusions is to not reheat to this temperature but then, since in all reasonable models $m_\varphi\sim\sqrt{F}$, it would no longer be possible for thermal effects to drive the theory to the metastable minimum \cite{Abel:2006cr, Craig:2006kx, Kaplunovsky:2007vd}.

Ultimately the task of building a viable model of metastable SUSY breaking now seems even more daunting.  Indeed, many existing models that appear to have long lived vacua are likely to be destabilised once Q-balls have been taken into account.  Fortunately all is not lost.  It is clear how SSR can be avoided, providing a concrete guide for future model building endeavours.

\phantom{woooo! spooky!}

\noindent\textbf{Acknowledgements:} I would like to thank Steven Abel for helpful comments and suggestions.  I would also like to thank Matthew McCullough and John March-Russell for stimulating discussions on this topic.  This work was supported by an STFC Postgraduate Studentship.

\providecommand{\href}[2]{#2}\begingroup\raggedright\endgroup

\end{document}